\documentclass[12pt,titlepage,oneside,showpacs]{revtex4}

\usepackage{amsmath}
\usepackage{amssymb}
\usepackage[dvips]{graphicx}
\usepackage{psfrag}
\usepackage{pstricks}
\usepackage{pst-node}
\usepackage{pst-plot}

 \psfrag{(a)}{(a)}
 \psfrag{(b)}{(b)}
 \psfrag{(c)}{(c)}
 \psfrag{(d)}{(d)}
 \psfrag{(e)}{(e)}
 \psfrag{(f)}{(f)}
 \psfrag{X}{$X$}
 \psfrag{Z}{$Z$}
 \psfrag{1}{$1$}

\newcommand{\qed}{\hspace*{\fill}$\square$}

 \newcommand{\sset}[1]{ \{#1\} }

 \newcommand{\prima}{^\prime}
 \newcommand{\primas}{^{\prime\prime}}

 \newcommand{\ket}[1]{|#1\rangle}
 \newcommand{\bra}[1]{\langle #1|}

\begin{document}

\title{An Interferometry-Free Scheme for \\Demonstrating Topological Order}

\author{H. Bombin and M.A. Martin-Delgado}
\affiliation{
Departamento de F\'{\i}sica Te\'orica I, Universidad Complutense,
28040. Madrid, Spain.
}

\begin{abstract}
We propose a protocol to demonstrate the topological order of a
spin-1/2 lattice model with four-body interactions. Unlike other
proposals, it does not rely on the controlled movement of
quasiparticles, thus eliminating the addressing, decoherence and
dynamical phase problems related to them. Rather, the protocol
profits from the degeneracy of the ground state. It involves the
addition of Zeeman terms to the original Hamiltonian that are used
to create holes and move them around in the system.
\end{abstract}


\maketitle

\section{introduction}

The notion of topological order (TO) has gradually become a new and
relevant topic in condensed matter physics \cite{wenbook04},
\cite{wenniu90}. It gives rise to a new paradigm of quantum phases
of matter which are endowed with long range correlations that cannot
be detected by local order parameters \cite{landau37},
\cite{ginzburg_landau50}. This is a new feature not associated with
the spontaneous breaking of a symmetry. Instead, the detection of
these new phases involve non-local order parameters that reflect the
global nature of these new highly strongly correlated systems.
Similarly, TOs turn out to be of great interest in quantum
information since they are considered as a resource of robustness
against the decoherence that typically affects all quantum systems
when we try to manipulate them with ease and control
\cite{kitaev97}. The possibilities range from quantum memories for
storage of quantum states \cite{dennis_etal02} to quantum computers
capable of performing a set of universal quantum operations
\cite{freedman_etal00a}, \cite{freedman_etal00b},
\cite{freedman_etal01}. The underlying mechanism for this robustness
arises in a typical scenario where the possible errors in the system
are local, while quantum logical operations are non-local and thus
potentially resilient to decoherence.

A practical way of describing a TO is as a strongly correlated
system with a quantum lattice Hamiltonian with the following
properties: i/ there is an energy gap between the ground state and
the excitations; ii/ the ground state is degenerate; iii/ this
degeneracy cannot be lifted by local perturbations. These features
reflects the topological nature of the system. In addition, a
signature of the TO is the dependence of that degeneracy on
topological invariants of the lattice where the system is defined,
like Betti numbers \cite{topo3D}. When the system is placed onto an
infinite plane, which has trivial topology, then the TO manifests
itself through the non trivial braiding properties of their
quasiparticle excitations \cite{levinwen05}: when two identical
particles are exchanged on the plane, their common wave function
picks up a nontrivial statistical phase. More generally, when one
particle completely encircles another particle, the state of the
system picks up a phase factor that is only trivial for bosons and
fermions, otherwise they are Abelian \cite{LM77}, \cite{wilczek82}
or non-Abelian anyons \cite{moore_read91}\cite{nano1}\cite{nano2}.
Thus, braiding statistics is also a signature of TO that can be
tried experimentally. Other signatures like the topological
entanglement entropy has also been proposed recently
\cite{kitaevpreskill06}, \cite{levinwen06}.

There has been a number of interesting experiments in order to
detect braiding statistics \cite{goldman_et_al05},
\cite{camino_et_al05}, \cite{camino_et_al07} in fractional quantum
Hall effect systems. This has turned out to be more elusive than
detecting fractional charge \cite{goldman_su95}. Thus, a number of
experimental proposals has been introduced aiming at providing
additional signatures of braiding statistics
\cite{dassarma_et_al05}, \cite{stern_halperin06},
\cite{bonderson_et_al06}, \cite{feldman_kitaev06},
\cite{law_feldman_gefen06} in fractional quantum Hall systems, both
Abelian and non-Abelian, which in turn would imply TO. For
non-Abelian gauge theories, it is also possible to detect anomalous
braiding statistics by interferometric means \cite{bais80},
\cite{Ogburn99}. There exist such intereferometric proposals for the
surface code introduced by Kitaev \cite{brennen},
\cite{han_et_al07}. This is the system in which we are interested
here.

In this paper we propose an alternative route to detect TO directly
and without having to resort to interferometry of quasiparticles to
probe their non-trivial braiding statistics. We use the fact that
the ground state degeneracy is sensitive to the topology of the
surface, which we can altere introducing Zeeman terms in certain
areas of the system. In particular, our scheme for detecting TO
relies on the notion of code deformations for surface codes
\cite{dennis_etal02}, \cite{raussendorf_et_al07},
\cite{CodeDeformation}.

%%%%%%%%%%%%%%%%%%%%%%%%%%%%%%%%%%%%
%%%%%%%%%%%%%%%%%%%%%%%%%%%%%%%%%%%%
\section{A model with string condensation}
%%%%%%%%%%%%%%%%%%%%%%%%%%%%%%%%%%%%
%%%%%%%%%%%%%%%%%%%%%%%%%%%%%%%%%%%%

\subsection{Hamiltonian and ground state}

\begin{figure}
 \psfrag{a}{$a$}
 \psfrag{b}{$b$}
 \psfrag{2}{2}
 \includegraphics[width=12cm]{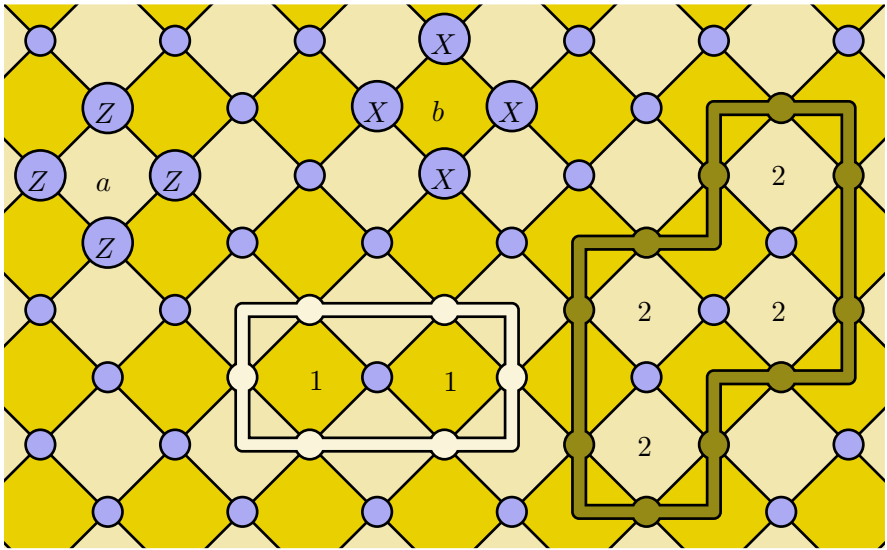}
 \caption
 {
Blue circles represent the spin-1/2 systems, lying on the sites of
the lattice. $Z^p$ ($X^p$) operators correspond to light (dark)
plaquettes like $a$ ($b$). The light (dark) string represents the
product of the plaquette operators of those dark (light) plaquettes
marked with a 1 (2).
 }
 \label{figura_system}
\end{figure}

The topologically ordered system that we consider here was
introduced by Kitaev \cite{kitaev97}. It is a 2-dimensional array of
spin-1/2 systems. Note that any subset $C$ of the spins can be
identified with a binary vector $(e_i)$, where $e_i=1$ if the $i$-th
spin belongs to $C$ and $e_i=0$ otherwise. Then, for each such set
$C$ we introduce the operators
\begin{equation}
X^C := \bigotimes_i \sigma_X^{e_i},\qquad Z^C := \bigotimes_i
\sigma_Z^{e_i}.
\end{equation}
Spins are located at the sites of a `chessboard' lattice, see
Fig.~\ref{figura_system}. The Hamiltonian is a sum of plaquette
operators $X^p$, $Z^p$ which depend on the coloring of the plaquette
$p$, dark or light,
\begin{equation}\label{Hamiltoniano}
H = - \sum_{p\in \mathcal P_D} g_p X^p -\sum_{p\in \mathcal P_L} g_p
Z^p,
\end{equation}
where $g_p > 0$ is the coupling constant at plaquette $p$, $\mathcal
P_D$ ($\mathcal P_L$) is the set of dark (light) plaquettes and we
identify each plaquette with the set of spins in its corners. The
spectrum of plaquette operators is $\sset{1,-1}$ and they commute,
so that the ground state is defined by the conditions
\begin{equation}\label{Condiciones}
X^p\ket\psi=Z^{p\prima}\ket\psi=\ket\psi, \qquad p\in \mathcal
P_D,\, p\prima\in \mathcal P_L,
\end{equation}
which must hold for all the plaquettes. If we consider that the
lattice extends to infinity or lyes on a sphere, there is no ground
state degeneracy. In particular, the unnormalized ground state takes
the form
\begin{equation}\label{Ground State}
\ket{\mathrm{GS}} = \prod_{p\in \mathcal P_D}(1+X^p)\ket{\psi_0},
\end{equation}
where $\psi_0$ is the state with all spins up. However, if the
topology of the surface is nontrivial the ground state is degenerate
\cite{kitaev97}.

\subsection{String operators}

A useful notion is that of dark and light strings, see
Fig.~\ref{figura_system} for examples. Light (dark) strings connect
light (dark) plaquettes, so that each string segment contains a
spin. Let $\gamma$ be a light string and $\gamma\prima$ a dark one.
Then we attach string operators to them, $X^\gamma$ and
$Z^{\gamma\prima}$, identifying strings with the sets of spins in
their segments. An important property is that $\sset {X^{\gamma},
Z^{\gamma\prima}}=0$ if $\gamma$ crosses $\gamma\prima$ and odd
number of times, $[X^{\gamma}, Z^{\gamma\prima}]=0$ otherwise.
Strings are either closed or have endpoints at plaquettes of their
color. When $\gamma$ and $\gamma\prima$ are closed we have
\begin{equation}\label{closed_strings}
[X^\gamma,H]=[Z^{\gamma\prima},H]=0.
\end{equation}
Among closed strings we find boundary strings, which receive this
name because they form the boundary of a portion of the surface.
Ground states can be characterized by the fact that if $\gamma$ and
$\gamma\prima$ are boundaries then
\begin{equation}\label{Condiciones_boundaries}
X^\gamma\ket\psi=Z^{\gamma\prima}\ket\psi=\ket\psi.
\end{equation}
This is equivalent to \eqref{Condiciones}, because plaquettes can be
identified with small boundaries, and boundary string operators are
products of plaquette operators. We can also rewrite \eqref{Ground
State} as
\begin{equation}\label{String Condensate}
\ket{\mathrm{GS}} = \sum_{\gamma\in\mathcal B^L} X^\gamma
\ket{\psi_0},
\end{equation}
where the elements of $\mathcal B_L$ are collections of boundary
strings. If we identify each state $X^\gamma\ket{\psi_0}$ with a
string configuration, that corresponding to $\gamma$, then the
ground state is a coherent superposition of string states. This is
why we say that the model is a string condensate \cite{levinwen05}.

\subsection{Excitations and topological charge}

The excitations of the system have a localized nature and are
subject to an energy gap. In particular, these quasiparticles are
related to plaquette operators, so that we say that the state $\ket
\psi$ has an excitation at plaquette $p$ if the corresponding
condition \eqref{Condiciones} is violated. The energy of the
quasiparticle is $\Delta = 2g_p$. Excited states can be obtained
from the ground state by applying open string operators: they create
quasiparticles at their endpoints.

\begin{figure}
 \psfrag{g}{$\gamma$}
 \psfrag{gp}{$\gamma\prima$}
 \psfrag{gpp}{$\gamma\primas$}
 \includegraphics[width=7cm]{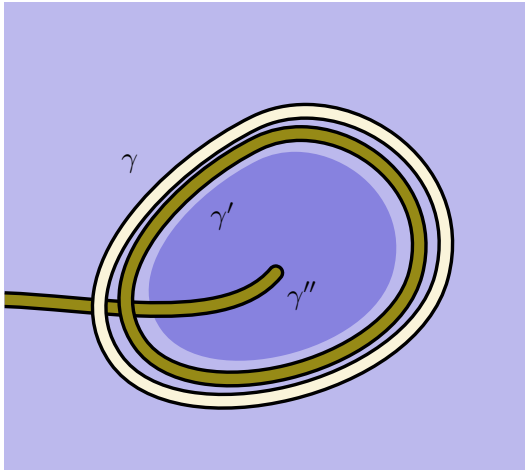}
 \caption
 {
  The total topological charge in the shaded region can be measured
  using the string operators $X^\gamma$ and $Z^{\gamma\prima}$.
  String operators from strings with endpoints in the region, such
  as $\gamma\prima$, change the charge of the region as they create or destroy a
  quasiparticle inside it.
 }
 \label{figura_carga}
\end{figure}

Excitations have a topological charge, which can be understood in
terms of string operators also. Suppose that we have several
excitations in the shaded region of Fig.~\ref{figura_carga}.
Consider a light string $\gamma$ and a dark string $\gamma\prima$
that surround the region. We construct four orthogonal projectors
that resolve the identity
\begin{equation}\label{Proyectores_carga}
P_{a,b} := \frac 1 4 \,(1+(-1)^a X^\gamma)\,(1+(-1)^b
Z^{\gamma\prima}), \qquad a,b=0,1.
\end{equation}
Each of the sectors $(a,b)$ projected by $P_{a,b}$ corresponds to a
different topological charge inside the region. These charges are
integrals of motion because of \eqref{closed_strings}. In the ground
state the charge is $(0,0)$, so this is the trivial charge. Consider
a dark string $\gamma\primas$ with an endpoint inside the region, as
in Fig.~\ref{figura_carga}. Then we have $Z^{\gamma\primas}P_{a,b} =
P_{a+1,b} Z^{\gamma\primas}$, with addition modulo two. Since
$\gamma\primas$ switches the excitations of the dark plaquettes in
its endpoints, we see that excitations of dark plaquettes carry the
charge $(1,0)$. Similarly, an excitation of a light plaquette
carries the charge $(0,1)$. It is easy to check that if a region is
divided on two subregions with charges $(a_1,b_1)$, $(a_2,b_2)$,
then its total charge is $(a_1+a_2, b_1+b_2)$, again with addition
modulo two. The topological nature of these charges relies in the
fact that when a charge $(a_1,b_1)$ is moved around a charge
$(a_2,b_2)$ the system will pick up a phase $(-1)^{a_1b_2+a_2b_1}$
which does not depend on the particular trajectory \cite{kitaev97}.

\section {Borders and topological degeneracy}

\subsection{Borders in surface codes}

The ground state subspace of the Hamiltonian \eqref{Hamiltoniano} is
a surface code, a kind of topological stabilizer code
\cite{kitaev97}. For our purposes here, a stabilizer code is a
subspace defined by certain conditions, which for surface codes are
\eqref{Condiciones_boundaries}. At first, surface codes were defined
in closed surfaces, but this has a limited use since it is difficult
to construct experimental setups with non-planar geometries.
However, even in the plane a nontrivial topology is possible if we
introduce borders \cite{bravyi_kitaev_bordes}, \cite{bmd_homologia}.

\begin{figure}
 \includegraphics[width=12cm]{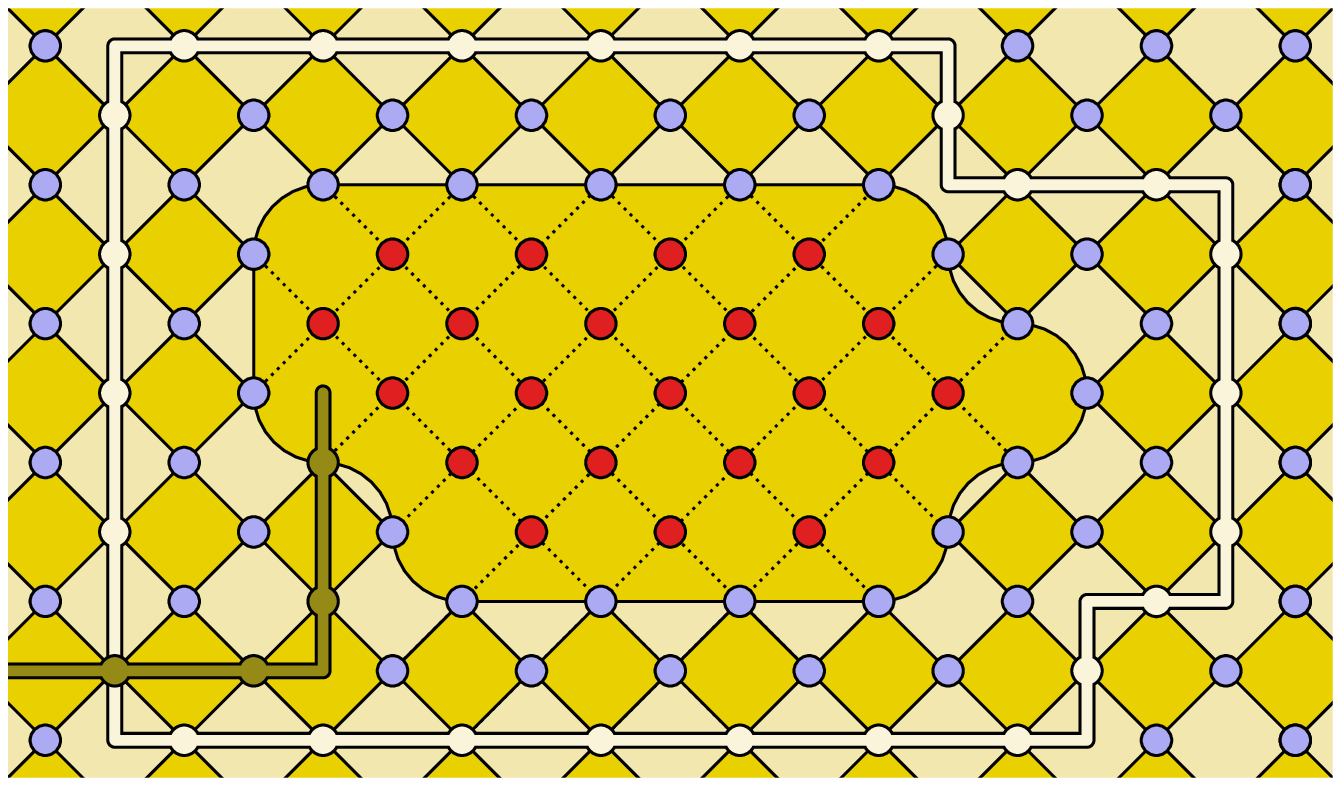}
 \caption
 {
This figure represents a dark hole in a surface code. Red sites
correspond to spins that are not part of the lattice. Note that a
dark hole is nothing but a missing big dark plaquette. The strings
on display are closed but not boundaries.
 }
 \label{figura_hole_code}
\end{figure}

Two kind of borders can be considered in surface codes, dark or
light. Borders change the concept of closed string. A dark (light)
string is closed either if it has no endpoints or if its endpoints
lye on dark (light) borders. Boundaries also change. A dark (light)
string is a boundary if it encloses a portion of surface which
contains no light (dark) borders. In surface codes, borders can be
introduced by changing the geometry of the lattice. In particular, a
dark (light) border corresponds to a missing big dark (light)
plaquette. Then the code can be described using the conditions
\eqref{Condiciones_boundaries} under the new notion of boundary
string. As an example, Fig.~\ref{figura_hole_code} shows a dark hole
in a lattice. It has been created by erasing several spins from the
lattice, shown in red, and rearranging the plaquettes accordingly.

The introduction of borders in surface codes allows to have
non-trivial topologies and thus a code subspace with dimension
greater than one. For example, if the surface is a disc with $h$
holes, with the borders of the same type, then the dimension of the
code is $2^h$\cite{bmd_homologia}. However, there is more to borders
than this. In particular, we can consider adding dynamics to the
picture. By changing the borders with time we can initialize,
transform and measure the states of the code \cite{CodeDeformation}.
This is a feature of surface codes that we would like to introduce
in the quantum Hamiltonian model, a possibility that we explore
next.

\subsection{Borders in the string condensate}

In principle, one could introduce borders in the quantum Hamiltonian
model \eqref{Hamiltoniano} simply by changing the geometry of the
lattice, that is, as in the surface code of
Fig.~\ref{figura_hole_code}. However, this would require the ability
to engineer a Hamiltonian in which for example a 3-body plaquette
term must exist next to a 4-body one and so on. Such a detailed
engineering is not feasible in many situations. Thus, we propose a
different setting, in which changes in the topology are produced by
modifying the original Hamiltonian through the introduction of
Zeeman terms and smooth spatial changes of the couplings.

\begin{figure}
 \psfrag{g1}{$\gamma_1$}
 \psfrag{g2}{$\gamma_2$}
 \psfrag{g1p}{$\gamma_1\prima$}
 \psfrag{g2p}{$\gamma_2\prima$}
 \psfrag{L}{$L$}
 \psfrag{Lb}{$L_B$}
 \psfrag{M}{$M$}
 \includegraphics[width=9cm]{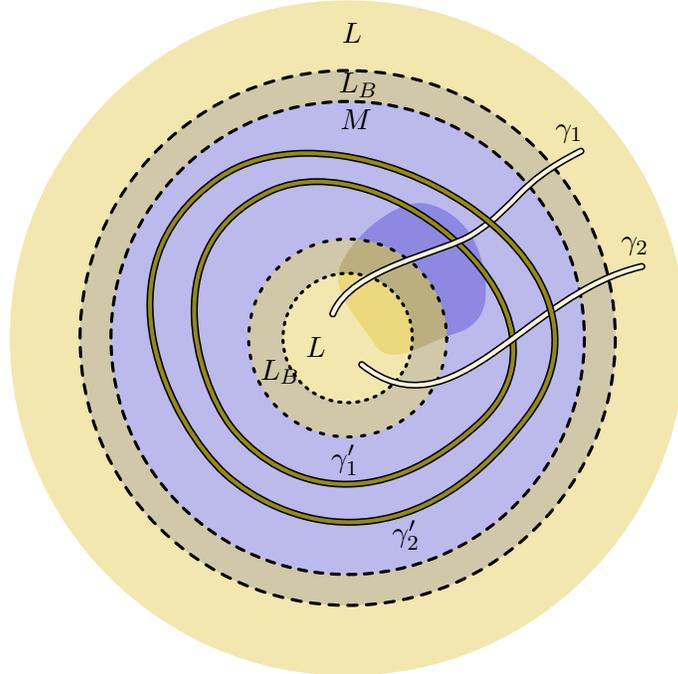}
 \caption
 {
 An example of how light borders are introduced in terms of the regions $L$,
 $L_B$ and $M$. In this case the main region $M$, in blue, has the topology of a disc with a hole, with
 both borders of light type. Examples of closed nontrivial strings
 are displayed. The $\gamma_i$ are closed because they have no endpoints. The $\gamma_i\prima$ are closed because their endpoints lye in $L$.
 The shaded area represents the support of a local operator,
 which neither encloses the interior $L$ region nor connects both $L$
 regions.
 }
 \label{figura_disco}
\end{figure}

We start by dividing the system surface in five regions, $M$, $D$,
$L$, $D_B$ and $L_B$. $M$ is the main system, where we are going to
keep the original Hamiltonian and thus the topological order remains
untouched. In the areas $D$ and $L$ there will be no topological
order. As for $D_B$ and $L_B$, these are thick boundaries that
separate $D$ from $M$ and $L$ from $M$, respectively. $D_B$ will
play the role of a dark boundary, and $L_B$ that of a light
boundary. An example can be seen in Fig.~\ref{figura_disco}, where
the geometry is that of a disc with a hole, with both borders of
light type.

We have to define the concepts of closed and boundary strings in our
new geometry with the five regions. A dark (light) string is closed
either if it has no endpoints or if they lye inside $D$ ($L$). A
dark (light) string is a boundary if it encloses a portion of
surface not containing any piece of $L\cup L_B$ ($D\cup D_B$). With
these definitions, we need a Hamiltonian that satisfies
\eqref{closed_strings} for closed strings and such that its ground
states satisfies \eqref{Condiciones_boundaries} for boundary strings
and there exists an energy gap to states not satisfying them. We
will first show why these conditions are enough to get the desired
properties, and afterwards give an example of a Hamiltonian that
satisfies the constrains.

We will work with a particular geometry to fix ideas, the disc with
a hole of Fig.\eqref{figura_disco}. Considering a general case has
no additional complications, but the discussion would be less
transparent. Let $V$ be the subspace defined by conditions
\eqref{Condiciones_boundaries}, so that the ground state subspace is
$V_{\mathrm {GS}}\subset V$. Consider the light string $\gamma_1$
and the dark string $\gamma\prima_1$ of Fig.\eqref{figura_disco}.
They are closed but not boundaries, and since they cross we have
$\sset{X^{\gamma_1},Z^{\gamma\prima_1}}=0$. These operators are the
$X$ and $Z$ operators of a qubit or two-level subsystem, both in $V$
and in $V_{\mathrm{GS}}$. Let us show this in detail. Note that
$X^{\gamma_1}$, $Z^{\gamma\prima_1}$ leave $V$ invariant, as closed
string operators always commute with boundary string operators. Then
we can choose an orthonormal basis $\sset{\ket {0; k}}_k$ for the
subspace of $V$ such that $Z^{\gamma\prima}=1$, which can be
completed in $V$ with the elements $\ket {1; k}:=X^{\gamma_1} \ket
{0; k}$, which satisfy $Z^{\gamma\prima_1}=-1$. In other words,
$V\simeq V\prima\otimes V_2$, with $V_2$ a two dimensional space.
The same is true for $V_{\mathrm{GS}}$, as follows from
\eqref{closed_strings}. That is, $V_{\mathrm{GS}}\simeq
V_{\mathrm{GS}}\prima\otimes V_2$ and $V\prima \simeq V\primas
\oplus V_{\mathrm{GS}}\prima$.

The point is that the degeneracy of the ground state that comes from
the qubit subsystem has a topological origin and cannot be lifted by
a small local perturbation. This a consequence of the fact that
there is an energy gap to states out of $V$ and that if $\sigma$ is
any local operator then
\begin{equation}\label{operador_local}
\bra{a;k}\sigma\ket{b;k\prima}=\delta_{a,b}\,
\bra{0;k}\sigma\ket{0;k\prima}, \qquad a,b=0,1
\end{equation}
We shall prove this equation in the following. For a local operator
we mean one with a support such as the shaded area in
Fig.\eqref{figura_disco}, which neither encloses the central $L$
region nor connects the interior and exterior $L$ regions. Then
there exist a light string $\gamma_2$ and a dark string
$\gamma\prima_2$, as in the figure, with the following properties.
First, they do not touch the support of $\sigma$, so that
$[X^{\gamma_2},\sigma]=[Z^{\gamma_2\prima},\sigma]=0$. Second, we
have the equivalences up to homology $\gamma_1\sim \gamma_2$,
$\gamma_1\prima\sim \gamma_2\prima$, so that
$X^{\gamma_1}X^{\gamma_2}=X^{\gamma_3}$ and
$Z^{\gamma_1\prima}Z^{\gamma_2\prima}=Z^{\gamma_3\prima}$ with
$\gamma_3$, $\gamma_3\prima$ boundaries. From these properties
\eqref{operador_local} follows immediately. This equation can also
be interpreted in terms of quantum error correction theory. It
states that we can correct information codified in the qubit
subsystem that has suffered a family of errors $\sset{E_i}$ as long
as any $\sigma=E_i^\dagger E_j$ is local \cite{operatorQEC}.

We now give an exactly solvable Hamiltonian that satisfies the
desired constrains. It takes the form
\begin{equation}\label{Hamiltoniano_2}
H = - \sum_{p\in \mathcal P_D} g_p\, X^p -\sum_{p\in \mathcal P_L}
g_p \,Z^p - \sum_i (\mu_i \,X^i + \nu_i\, Z^i),
\end{equation}
where $i$ runs over the sites of the lattice, $g_p, \mu_i, \nu_i
\geq 0$ are coupling constants and we identify a site $i$ with the
set $\sset{i}$. As long as $\nu_i=0$ ($\mu_i=0$) for all the sites
$i$ that lye on the corner of a dark (light) plaquette $p$ with
$g_p>0$, the Hamiltonian is exactly solvable because all the
non-vanishing terms are commuting projectors. Then, the ground state
subspace is characterized by the conditions
\begin{equation}\label{Condiciones_2}
X^p\ket\psi=Z^{p\prima}\ket\psi=Z^i\ket\psi=X^j\ket\psi=\ket\psi,
\end{equation}
which must hold for all the dark plaquettes $p$ with $g_p>0$, light
plaquettes $p\prima$ with $g_{p\prima}>0$, sites $i$ with $\nu_i>0$
and sites $j$ with $\mu_j>0$. It is possible to choose the couplings
in such a way that the conditions
(\ref{closed_strings},\ref{Condiciones_boundaries}) are satisfied.
In particular, $\mu_i>0$ ($\nu_i>0$) must be fulfilled in $L$ ($D$),
whereas $\mu_i=0$ ($\nu_i=0$) in $M\cup D\cup D_B$ ($M\cup L\cup
L_B$). Also, $g_p>0$ must hold for dark (light) plaquettes in $M\cup
L_B$ ($M\cup D_B$), whereas $g_p=0$ in $D$ ($L$). All this can be
done in such a way that the couplings vary smoothly across the
surface, due to the thickness of the boundary regions $L_B$ and
$D_B$.

As a result of the above construction, we will find in general a
local degeneracy in the ground state, since there exist areas in
$L_B$ ($D_B$) where the only non-zero coupling is $g_p$ in dark
(light) plaquettes. This local degeneracy can be removed by letting
the support of $\mu_i$ ($\nu_i$) overlap with that of the $g_p$ of
light (dark) plaquettes. In doing so, the Hamiltonian is no longer
exactly solvable, but it will fulfill the required conditions at
least as long as the overlap is not too big. To see this, note that
we can write the Hamiltonian as $H\prima = H + H_p$, where $H_p$
contains those terms that do not commute with all the terms of
$H\prima$. Then $H$ is exactly solvable and has the required
properties. Also, $[H_p,H]=0$. Indeed, each of the terms of $H_p$
commutes with each of the terms in $H$. Thus, an small $H_p$ will
not destroy the properties of $H$ discussed above. Still, if the
overlap is too big, a level crossing could occur taking the ground
state out of the subspace $V$ described by
\eqref{Condiciones_boundaries}.

%%%%%%%%%%%%%%%%%%%%%%%%%%%%%%%%%%%%
%%%%%%%%%%%%%%%%%%%%%%%%%%%%%%%%%%%%
\subsection{Surface deformation}\label{seccion_code_deformation}
%%%%%%%%%%%%%%%%%%%%%%%%%%%%%%%%%%%%
%%%%%%%%%%%%%%%%%%%%%%%%%%%%%%%%%%%%

Once one is able to engineer the Hamiltonian \eqref{Hamiltoniano_2},
the next step is to adiabatically modify the couplings so that the
geometry of the surface changes slowly with time. Here we can
distinguish too kind of such surface deformations. First, we can
perform deformations in which only the geometry of the surface, not
its topology, change with time. When the initial and the final state
of the system are the same, these produce a continuous map of the
surface onto itself, so that in particular strings get transformed.
This gives a string operator mapping, which amounts to perform a
definite operation on the encoded subsystem \cite{CodeDeformation}.
Second, deformations that change the topology can be considered,
such as introducing or destroying holes and cutting or gluing pieces
of the main surface $M$. These kind of processes change the
topological degeneracy of the ground state. When it grows, the new
degrees of freedom will be initialized in a definite way
\cite{CodeDeformation}, due to topological considerations. When it
decreases, the lost degrees of freedom get mapped to possible
excitations in the final state.

\begin{figure}
 \psfrag{g1}{$\gamma_1$}
 \psfrag{g2}{$\gamma_2$}
 \psfrag{g3}{$\gamma_3$}
 \psfrag{L}{$D$}
 \psfrag{Lb}{$D_B$}
 \psfrag{M}{$M$}
 \includegraphics[width=12cm]{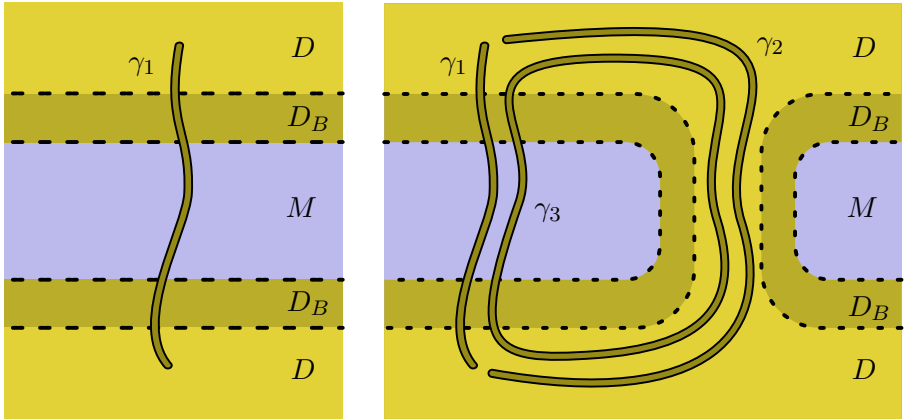}
 \caption
 {
A deformation in which two separated $D$ regions are put together,
which amounts to cut the main region $M$. To the left, the geometry
before the cut is done. We suppose that $\gamma_1$ is a nontrivial
closed string. To the right, the geometry after the cut. Now
$\gamma_1$ is a boundary string, and so are $\gamma_2$ and
$\gamma_3$. If $Z^{\gamma_2}=1$, then $Z^{\gamma_1}=Z^{\gamma_3}$,
that is, the cut maps the value of $Z^{\gamma_1}$ to the light
plaquete charge in the region surrounded by $\gamma_3$.}
\label{figura_medida}
\end{figure}

This deserves a more detailed explanation. Consider for example the
surface deformation illustrated in Fig.~\eqref{figura_medida}, where
too separate pieces of region $D$ get connected, producing a cut in
$M$. Consider the dark string $\gamma$ that connects both $D$ areas.
We want to show that the deformation amounts to a measurement of
$Z^\gamma$. Before the deformation $\gamma$ is closed --- and we
assume that nontrivial --- and after the deformation it is a
boundary. Because of the local nature of the deformation, it cannot
change the value of $Z^{\gamma_1}$, which lyes outside the area
where the action occurs. But if $Z^{\gamma_1}=-1$, then the final
state cannot fulfill conditions \eqref{Condiciones_boundaries} and
thus it is not a ground state. Which excitations should we find? To
answer this, let us suppose that the coupling $\mu_i$ is big enough
in $D$, so that in the final state we know that $Z^{\gamma_2}=1$ is
fulfilled for any dark string $\gamma_2$ lying inside $D$. Then for
the dark boundary string $\gamma_3$ formed by composing $\gamma_1$
and $\gamma_2$, see Fig.~\ref{figura_medida}, and for the final
state $\ket \psi$ we have
$Z^{\gamma_3}\ket\psi=Z^{\gamma_1}Z^{\gamma_2}\ket\psi=Z^{\gamma_1}\ket\psi$.
Since the value $Z^{\gamma_3}=\pm 1$ is related to light plaquette
charge inside $\gamma_3$ through \eqref{Proyectores_carga}. We see
that the cutting process, as announced, amounts to a measurement of
$Z^\gamma_2$, as its value is mapped to the possible appearance of
charge at both sides of the cut.

For the previous analysis, the deformation needs not really be
adiabatic. It is enough if we can guarantee that there are no
excitations inside $D$. The particularity of the adiabatic case is
that we expect to find a final state with a single light plaquette
excitation at each side of the cut, since this is a state in a local
energy minimum. We will see an application of these measurements
through surface cutting --- and indeed of all the mentioned kinds of
surface deformations --- in the scheme to demonstrate the
topological character of the phase discussed below .

It is worth mentioning that these ideas can be used to adiabatically
initialize the topologically ordered phase. In this regard, a
question was raised in \cite{Adiabatic Initialization} about how to
adiabatically initialize these systems so that the topological
protection is present all along the way and not only after reaching
the topological phase. The answer is that, instead of initializing
the whole system at a time, one should progressively grew it from a
small island till the desired surface is covered. In surfaces with
non-trivial topology, this means that at some point two different
borders of the system will fuse. At that point the degeneracy of the
ground state will change, as new nontrivial string operators appear.
The eigenvalues of the new string operators that run along such
junctions are necessarily one \cite{CodeDeformation}, and thus the
final particular ground state of the system is perfectly determined.

%%%%%%%%%%%%%%%%%%%%%%%%%%%%%%%%%%%%
%%%%%%%%%%%%%%%%%%%%%%%%%%%%%%%%%%%%
\section{A scheme for demonstrating topological order}\label{secc_esquema}
%%%%%%%%%%%%%%%%%%%%%%%%%%%%%%%%%%%%
%%%%%%%%%%%%%%%%%%%%%%%%%%%%%%%%%%%%

When trying to demonstrate TO, the usual approaches focus on
interferometric experiments with quasiparticles in which
topologically different paths are compared. An immediate problem of
such approaches is that the required quasiparticle superposition of
states are subject to decoherence due to their localized nature and
the presence of a noisy environment. Also dynamical phases have to
be taken into account and properly controlled. Here we adopt a
different approach that eliminates both problems by focusing on the
ground state degeneracy. The idea is to show that the outcome of
certain processes depends only on topological properties, thus
revealing the topological nature of the system.

\begin{figure}
 \psfrag{g1}{$\gamma_1$}
 \psfrag{g2}{$\gamma_2$}
 \psfrag{g1p}{$\gamma_1\prima$}
 \psfrag{g2p}{$\gamma_2\prima$}
 \psfrag{g3p}{$\gamma_3\prima$}
 \includegraphics[width=12cm]{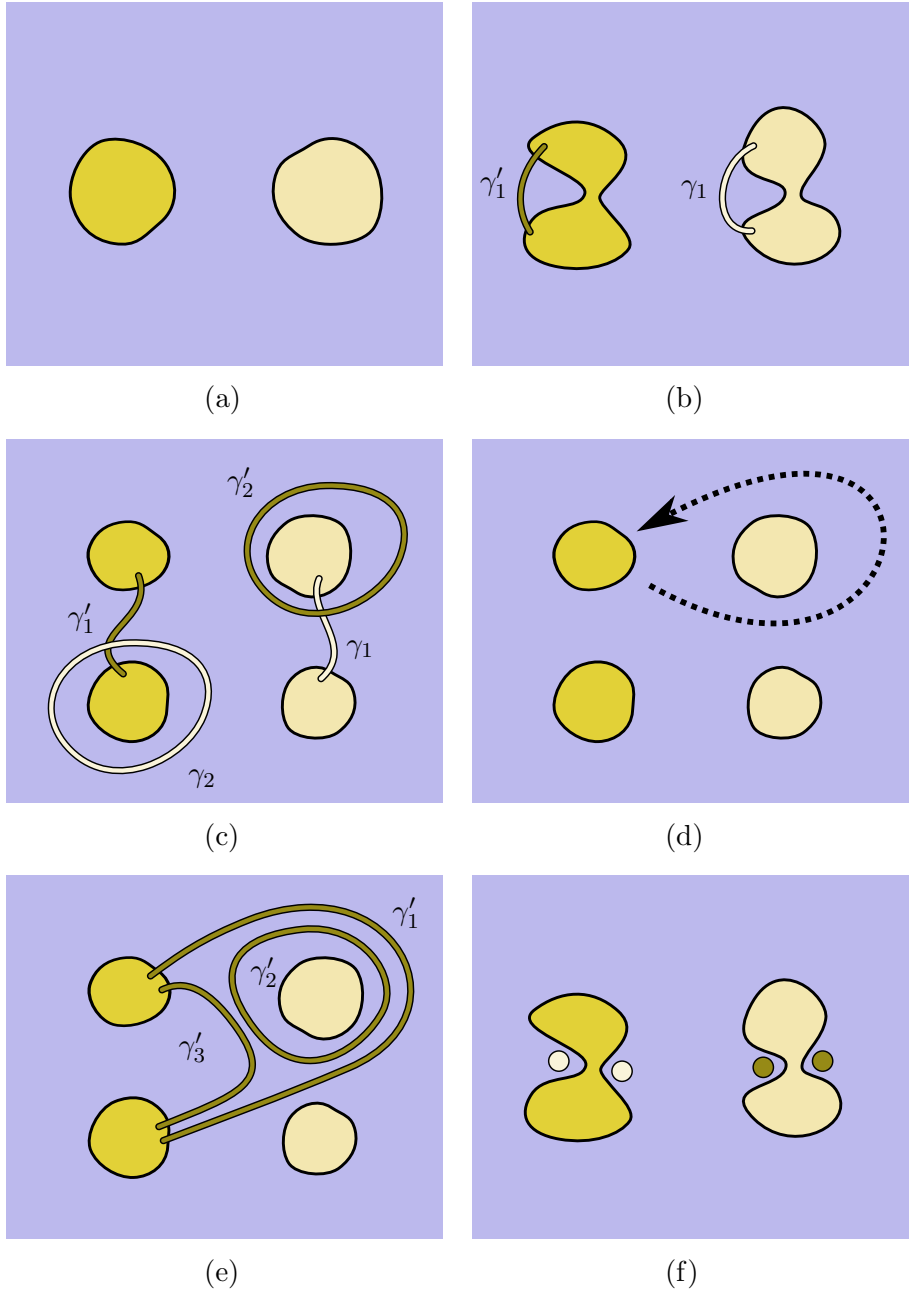}
 \caption
 {
 A step-by-step representation of the proposed scheme, as explained in section \ref{secc_esquema}.}
 \label{figura_scheme}
\end{figure}

The scheme is as follows. We start by making a pair of holes in our
system, a dark one and a light one, see Fig.~\ref{figura_scheme}(a).
Then we deform both of them as in Fig.~\ref{figura_scheme}(b), till
they are separated into two pieces. Notice that since in figure
Fig.~\ref{figura_scheme}(b) $\gamma_1$ and $\gamma_1\prima$ are
boundaries we have $X^{\gamma_1}=Z^{\gamma_1\prima}=1$. After the
hole breaks into two pieces they still must have the same value
because it is a global property\cite{CodeDeformation}, so that we
reach the situation in Fig.~\ref{figura_scheme}(c), where
$X^{\gamma_2}$ and $Z^{\gamma_2\prima}$ have completely undefined
values since
$\sset{X^{\gamma_1},Z^{\gamma_2\prima}}=\sset{X^{\gamma_2},Z^{\gamma_1\prima}}=0$.
We then proceed to move one of the dark holes along a closed path.
Suppose for the moment that the path is as the one shown in
Fig.~\ref{figura_scheme}(d), that is, that it encloses one of the
light holes. The point is that, after this has been accomplished,
the string operators have deformed accordingly. For example,
$\gamma_1\prima$ has changed and now its place is occupied by
$\gamma_3\prima$, see Fig.~\ref{figura_scheme}(e). If $\ket\psi$ is
the state corresponding to that figure, we have
$Z^{\gamma_3\prima}\ket\psi=Z^{\gamma_1\prima}Z^{\gamma_2\prima}\ket\psi
=Z^{\gamma_2\prima}\ket\psi $. A similar analysis holds for a light
string connecting the light holes. When we finally refuse the holes,
as in Fig.~\ref{figura_scheme}(f), we are measuring these string
operators that connect each pair of holes, which have a completely
undefined value, so that there exists a 1/2 probability that we find
charges at both sides of the fusion point, as follows from the
explanation in section \ref{seccion_code_deformation}. The problem
of how to detect this charge would depend on the particular
experimental situation.

Now return to the path in Fig.~\ref{figura_scheme}(d) and consider
any line $l$ joining both light holes. We can imagine many other
closed paths, some of them never crossing this line and others
crossing it many times. The topological property in which we are
interested is the number of times a path crosses $l$. If the number
is odd, the situation is the one described above. If it is even,
then it is equivalent to doing nothing and if we refuse the holes we
will never find charges \cite{CodeDeformation}. Also, note that
several quasiparticles could be created during the fusion of the
holes if it is not adiabatic, but the evenness or oddness of the
number of particles created at each side is topologically protected
since it gives the total topological charge.

Thus the topological nature of the system manifests in the fact that
the experiment is sensitive to the topology of the chosen path.
Moreover, the underlying $Z_2$ nature of the system is revealed
also: only the evenness or oddness of the linking number is
important. With this scheme we have introduced a new way to probe
the existence of a TO. It  does not involve the ability to
manipulate individual quasiparticle excitations, but instead relies
solely on the peculiar ground state properties of topologically
ordered quantum systems.

%%%%%%%%%%%%%%%%%%%%%%%%%%%%%%%%%%%%
%%%%%%%%%%%%%%%%%%%%%%%%%%%%%%%%%%%%
\section{Final remarks}
%%%%%%%%%%%%%%%%%%%%%%%%%%%%%%%%%%%%
%%%%%%%%%%%%%%%%%%%%%%%%%%%%%%%%%%%%

Although we have restricted ourselves to the Kitaev $Z_2$ model, it
is possible to consider generalizations to $Z_D$ systems or even
non-abelian models. In this regard, as noted above, the definition
of holes in terms of open strings is a natural starting point and  a
much more richer family of 'holes' is expected in such systems, but
the basic mechanism for topological detection without resorting to
quasiparticle interferometry remains the same.

\noindent {\em Acknowledgements} We acknowledge financial support
from a PFI fellowship of the EJ-GV (H.B.), DGS grant  under contract
BFM 2003-05316-C02-01 (M.A.MD.), and CAM-UCM grant under ref.
910758.

\end{document}